%% file: Dielectric_constant_resubmission.tex
\documentclass[english,aps,article,prx,reprint,microtype,longbibliography,superscriptaddress]{revtex4-1}
\usepackage[LGR,T1]{fontenc}
\usepackage{color}
\usepackage{babel}
\usepackage{booktabs}
\usepackage{amssymb}
\usepackage{graphicx}
\usepackage{subscript}
\usepackage[unicode=true,
 bookmarks=true,bookmarksnumbered=false,bookmarksopen=false,
 breaklinks=false,pdfborder={0 0 0},pdfborderstyle={},backref=false,colorlinks=true]
 {hyperref}
\hypersetup{
 pdfauthor={Ambroise van Roekeghem}}

\AtBeginDocument{%
    \newwrite\bibnotes
    \def\bibnotesext{Notes.bib}
    \immediate\openout\bibnotes=\jobname\bibnotesext
    \immediate\write\bibnotes{@CONTROL{REVTEX41Control}}
    \immediate\write\bibnotes{@CONTROL{%
    apsrev41Control,author="08",editor="1",pages="1",title="0",year="1"}}
     \if@filesw
     \immediate\write\@auxout{\string\citation{apsrev41Control}}%
    \fi
}%

\makeatletter

\DeclareRobustCommand{\greektext}{%
  \fontencoding{LGR}\selectfont\def\encodingdefault{LGR}}
\DeclareRobustCommand{\textgreek}[1]{\leavevmode{\greektext #1}}
\ProvideTextCommand{\~}{LGR}[1]{\char126#1}

\providecommand{\tabularnewline}{\\}

\@ifundefined{date}{}{\date{}}
\date{}

\makeatother

\begin{document}
\title{High-throughput study of the static dielectric constant at high-temperatures
in oxide and fluoride cubic perovskites}
\author{Ambroise van Roekeghem}
\affiliation{Universit\'e Grenoble Alpes, CEA, LITEN, 17 Rue des Martyrs, 38054 Grenoble, France}
\email{ambroise.vanroekeghem@cea.fr}

\author{Jes\'{u}s Carrete}
\affiliation{Institute of Materials Chemistry, TU Wien, A-1060 Vienna, Austria}
\author{Stefano Curtarolo}
\affiliation{Center for Autonomous Materials Design, Duke University, Durham, NC 27708,
USA}
\affiliation{Department of Mechanical Engineering and Materials Science, Duke University, Durham, NC 27708, USA}
\author{Natalio Mingo}
\affiliation{Universit\'e Grenoble Alpes, CEA, LITEN, 17 Rue des Martyrs, 38054 Grenoble, France}
\date{\today}
\begin{abstract}
Using finite-temperature phonon calculations and the Lyddane-Sachs-Teller
relations, we calculate ab initio the static dielectric constants of
78 semiconducting oxides and fluorides with cubic perovskite structures
at 1000\,K. We first compare our method with experimental measurements,
and we find that it succeeds in describing the temperature dependence
and the relative ordering of the static dielectric constant $\epsilon_{DC}$
in the series of oxides BaTiO$_{3}$, SrTiO$_{3}$, KTaO$_{3}$. We show that the effects of
anharmonicity on the ion-clamped dielectric constant, on Born charges, and on phonon lifetimes, 
can be neglected in the framework of our high-throughput study. 
Based on the high-temperature phonon spectra, we find that the
dispersion of $\epsilon_{DC}$ is one order of magnitude larger amongst oxides than fluorides
at 1000\,K. We display the correlograms of the dielectric constants with simple structural descriptors,
and we point out that $\epsilon_{DC}$ is actually well correlated
with the infinite-frequency dielectric constant $\epsilon_{\infty}$,
even in those materials with phase transitions in which $\epsilon_{DC}$
is strongly temperature-dependent.
\end{abstract}
\maketitle

\section{Introduction}

\label{sec:Introduction}

Capacitors for emerging power electronics -- including those in avionics and automotive systems -- must operate in specific temperature ranges, often outside ambient conditions. Perovskite materials have been widely used as ceramic dielectrics in capacitors, exhibiting different capacitances as a function of temperature. For example, BaTiO$_{3}$ has been used for ambient-condition applications since it has a Curie point close to room temperature, yielding high static dielectric constants. However, at higher temperatures, its lattice properties change, and capacitance drops. To optimize ceramic dielectrics for custom temperature ranges, it would be useful to perform high-throughput searches at finite temperatures. The static dielectric constant can be computed from first principles, {\it e.g.} using density functional perturbation
theory (DFPT) \citep{Gonze_DFPT_dielectric}. Such calculations have
already been performed in a high-throughput fashion for
a large number of materials \citep{Petousis_HT_dielectric_constant,Petousis_scientific_data_dielectric}.
Still, the approach is limited to ground-state properties, irrelevant for strongly anharmonic materials like perovskites,
whose phonon spectrum and crystal structure vary significantly with
temperature. In previous studies, this temperature dependence has
been obtained via the electro-optic coefficients, which link the polarization
with the refractive index \citep{DiDomenico_Wemple_Electro-optical,Burns_Scott_refraction_ferroelectrics,Burns_Dacol_PhysRevB.28.2527,Bernasconi_electro-optic,Veithen_Ghosez_PhysRevB.71.132101}.
The polarization can then be computed using effective Hamiltonians
\citep{Zhong_effective_Hamiltonian,Veithen_Ghosez_PhysRevB.71.132101}
or with molecular dynamics \citep{Silvestrelli_MD_infrared,Pagliai_MD_infrared}.

Here, we use finite-temperature lattice dynamics calculations
to compute the static dielectric constant of 78 oxide and fluoride
cubic perovskites at high temperature, directly from the temperature-dependent
phonon spectra. Our high-throughput methodology is particularly
interesting for the search of new materials that could operate under
specific conditions. In particular, with the recent development of
sensors and electronics designed to operate close to the engine of
airplanes or vehicles, the industry would benefit from compounds that
are more adapted to high-temperature conditions, especially for use
in capacitors \citep{Zeb_high_T_dielectrics}.

We first describe the method, which is based on the Lyddane-Sachs-Teller (LST)
relations \citep{LST_relations,LST_Cochran} combined with temperature-dependent
lattice dynamics calculations \citep{Ambroise_ScF3}, and demonstrate
its reliability by computing the dielectric constants of BaTiO$_{3}$,
SrTiO$_{3}$ and KTaO$_{3}$. We also evaluate the impact of the modifications
of the Born charges and of the electronic dielectric constant due
to thermal effects on the value of the static dielectric constant,
as well as the effect of lifetime broadening, in the case of SrTiO$_{3}$.
In a second part, we build on the high-throughput finite-temperature
phonon spectra that we obtained in our previous work \citep{Ambroise_perovskites_kappa}
to study the dielectric constant of oxide and fluoride cubic perovskites
at high temperature. We find that oxides present a larger diversity
of values, and that the ion-clamped value of the dielectric constant is unexpectedly
well correlated with the high-temperature dielectric properties.

\section{Ab initio calculations of the static dielectric constant at finite
temperature}

\label{sec:Finite-T calculations}

The ionic contribution to the relative permittivity can be obtained
from the generalized LST relations \citep{LST_Cochran}.
The crystal is isotropic in cubic perovskites, leading to 
\begin{equation}
\frac{\epsilon_{DC}}{\epsilon_{\infty}}=\prod_{j}\left(\frac{\omega_{Lj}}{\omega_{Tj}}\right)^{2},\label{eq:LST}
\end{equation}
with $\epsilon_{DC}$ the static dielectric constant, $\epsilon_{\infty}$
the ion-clamped dielectric constant, and $\omega_{L}$ and $\omega_{T}$
the long-wavelength longitudinal and transverse optical frequencies.
The splitting between longitudinal (LO) and transverse (TO) modes, along with
$\epsilon_{\infty}$, can be obtained from DFPT as implemented for
instance in VASP \citep{Kresse_DFPT}. To obtain effective second-
order interatomic force constants at finite temperature including
anharmonic effects, we use the method presented in Ref.\,\citep{Ambroise_ScF3},
which uses a regression analysis of forces from density functional
theory coupled with a harmonic model of the quantum canonical ensemble.
This is done in an iterative way to achieve self-consistency of the
phonon spectrum. The value of $\epsilon_{\infty}$ is taken from ground-state
calculations (see discussion below). LO/TO splitting is obtained from the non-analytical correction for $q\rightarrow0$
using ground-state Born charges, as implemented in phonopy \citep{NAC_Wang**3,phonopy}.
In practice, we adiabatically turn on the non-analytical correction
in order to track the different pairs of modes that are relevant in
Eq. (\ref{eq:LST}).

We first perform the calculations for a series of three oxides, BaTiO$_{3}$
(at 400\,K, 700\,K and 1000\,K), SrTiO$_{3}$ (at 300\,K, 500\,K
and 1000\,K) and KTaO$_{3}$ (at 100\,K, 500\,K and 1000\,K, including
also spin-orbit coupling), using the PBEsol functional, a 4x4x4 supercell
and a 30x30x30 phonon wavevector grid. For other temperatures, we
simply interpolate the force constants linearly: under the approximation
that the eigenvectors are unchanged, the soft mode would then follow
a Curie law,
\begin{equation}
\omega^{2}=A(T-T_{c}).
\end{equation}
This interpolation also allows us to estimate the Curie temperature
of those compounds. We obtain for BaTiO$_{3}$: $T_{c}\thickapprox$
270\,K, for SrTiO$_{3}$: $T_{c}\thickapprox$ 140\,K and for KTaO$_{3}$:
$T_{c}\thickapprox$ 80\,K. This difference of about 100\,K compared
to the experimental values corresponds to very tiny energy scales (less than 10 meV),
and is comparable to what has been found in other studies \citep{Tadano_SCPH}.

\begin{figure}[h]
\begin{centering}
\includegraphics[width=8.5cm]{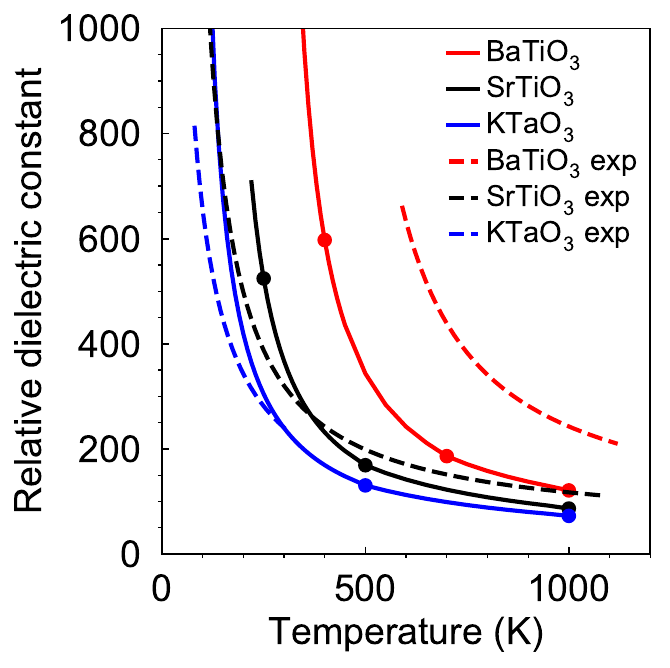}
\par\end{centering}
\caption{A comparison between the static dielectric constants of BaTiO$_{3}$
(red), SrTiO$_{3}$ (black) and KTaO$_{3}$ (blue), as calculated
ab initio (dots, the value at 100\,K for KTaO$_{3}$ is 1794, out
of the scale) and using a linear interpolation of the interatomic
force constants for other temperatures (full lines), and the fitted
expressions $\epsilon=\epsilon_{L}+C/(T-T_{c})$ obtained from experimental
measurements in Rupprecht and Bell, Phys. Rev. B 135, A748 (1964)
\citep{Rupprecht_dielectric_perovskites} (dashed lines).\label{fig:STO_BTO_KTO}}
\end{figure}

Our results for the static dielectric constant are displayed in Fig.
\ref{fig:STO_BTO_KTO} and compared with the experimental data from
Rupprecht and Bell \citep{Rupprecht_dielectric_perovskites}. As expected
from the transition temperatures, there is a shift of the curve of
about 150\,K for BaTiO$_{3}$ compared to the experiment,
and a difference in the temperature dependence, reflecting the limitations
of the method. Notably, this kind of calculations can be sensitive
to the choice of the exchange and correlation functional, already for
the ground-state properties \citep{Petousis_HT_dielectric_constant}.
Nevertheless, the order of magnitude and the respective ordering of
the values are correct.

The Born charges and electronic dielectric constant $\epsilon_{\infty}$
are also modified by temperature, due to thermal expansion and to
ionic displacements \citep{Ghosez_Michenaud_Gonze_PhysRevB.58.6224,Veithen_Ghosez_PhysRevB.71.132101}.
Furthermore, the LST relations can be generalized to account, for
instance, for the effect of lifetime broadening \citep{Barker_PhysRevB.12.4071,Noh_Sievers_PhysRevLett.63.1800}.
We now evaluate the impact of those effects on the value of the static
dielectric constant, in the case of SrTiO$_{3}$. Table\,\ref{tab:Born} show the computed
thermal average of the Born charges and of $\epsilon_{\infty}$
on 3 different configurations of our 4x4x4 supercell with a 2x2x2
grid.
We observed that $\epsilon_{\infty}$ is decreased by anharmonic effects,
in agreement with Ref.\,\citep{Veithen_Ghosez_PhysRevB.71.132101},
whereas thermal expansion actually increases $\epsilon_{\infty}$. If
we now include these effects into $\epsilon_{DC}$ using the corresponding
non-analytical correction, we find that at 1000\,K $\epsilon_{DC}$
is corrected to 76.6 instead of 86.5. The main correction comes from
the difference in the splitting of the highest optical mode due to
the thermally averaged Born charges. For the generalized LST relations,
we evaluate the expression of Barker \citep{Barker_PhysRevB.12.4071}:
\begin{equation}
\frac{\epsilon_{DC}}{\epsilon_{\infty}}=\prod_{j}\left(\frac{\left|\omega_{Lj}\right|}{\omega_{Tj}}\right)^{2},\label{eq:LST-Barker}
\end{equation}
where $\omega_{L}=\Re\omega_{L}-i\gamma/2$. We use the previously
calculated value as the real part, and $\gamma$ is evaluated from
the lifetime of the corresponding phonon mode computed at the $\Gamma$
point with almaBTE \citep{almaBTE_CompPhysComm2017} on a 15x15x15
phonon wavevector grid, using the temperature-dependent 2nd and 3rd-order
force constants. Table\,\ref{tab:Lifetime} shows the real and
imaginary parts for all LO modes in SrTiO$_{3}$. This translates
into a correction of the static dielectric constant from 86.5 to 87.0
at 1000\,K. Thus those two effects account for a modification of
about 10\% of $\epsilon_{DC}$, which is similar to the
potential error in the estimation of the phonon frequencies and of
the transition temperature, in particular if we compare between different
functionals. This picture might be different close to
displacive phase transitions, where lifetimes can typically become
shorter and ionic fluctuations more important. Since we are here mostly
interested in high-temperature properties and in global trends over
the whole family of fluoride and oxide cubic perovskites, we choose
to neglect those effects in the rest of the article.

\begin{table}
\begin{tabular}{cccccc}
\hline 
 & $\epsilon_{\infty}$ & $Z_{Sr}^{*}$ & $Z_{Ti}^{*}$ & $Z_{O_{\perp}}^{*}$ & $Z_{O_{||}}^{*}$\tabularnewline
\hline 
Ground state & 6.36 & 2.55 & 7.35 & -2.04 & -5.82\tabularnewline
1000\,K & 6.27 & 2.54 & 6.76 & -1.94 & -5.41\tabularnewline
\hline 
\end{tabular}

\caption{Thermal average of $\epsilon_{\infty}$ and of the Born effective
charges in SrTiO$_{3}$ at 1000\,K.\label{tab:Born}}

\end{table}

\begin{table}
\begin{tabular}{cccccccccc}
\hline 
 & $\Re\omega_{L_{1}}$ & $\tau_{L_{1}}$ & $\Im\omega_{L_{1}}$ & $\Re\omega_{L_{2}}$ & $\tau_{L_{2}}$ & $\Im\omega_{L_{2}}$ & $\Re\omega_{L_{3}}$ & $\tau_{L_{3}}$ & $\Im\omega_{L_{3}}$\tabularnewline
1000\,K & 32.4 & 0.30 & 1.7 & 84.2 & 0.11 & 4.4 & 145.2 & 0.15 & 3.4\tabularnewline
\hline 
\end{tabular}

\caption{Real part (rad/ps), lifetime (ps) and imaginary part (rad/ps) of the
LO modes in SrTiO$_{3}$ at 1000\,K.\label{tab:Lifetime}}
\end{table}

\section{High-throughput results}

We now apply the same methodology to the set of 92 oxide and fluoride
perovskites that we identified as mechanically stable at 1000\,K
in our previous work \citep{Ambroise_perovskites_kappa}. Because
of the reduced precision (notably, a smaller 2x2x2 supercell) and
of the high-throughput methodology, the absolute values are expected
to be less accurate. Still, the general trends can give useful physical
insight. We have recomputed the phonon spectra of all compounds using
the latest version of our code, which has been made public recently
\citep{QSCAILD}. We also used the PBEsol functional that is more
adapted to most of these compounds, except from the ones including
Fe that were then found to be metals and excluded from our set. Spin-orbit
coupling was also taken into account for all compounds containing
atoms heavier than Hf. This gave us a new total of 78 oxide and fluoride
perovskites mechanically stable in the cubic phase at 1000\,K.

\begin{table*}
\begin{centering}
\begin{tabular}{ccccccccccccccccccccccccccccccc}
\toprule 
 & $\epsilon_{1000}$ & $\epsilon_{300}$ &  &  &  &  &  & $\epsilon_{1000}$ & $\epsilon_{300}$ &  &  &  &  &  & $\epsilon_{1000}$ & $\epsilon_{300}$ &  &  &  &  &  & $\epsilon_{1000}$ & $\epsilon_{300}$ &  &  &  &  &  & $\epsilon_{1000}$ & $\epsilon_{300}$\tabularnewline
\textcolor{red}{TlNbO$_{3}$} & 190 & 2954 &  &  &  &  & BeScF$_{3}$ & 58 & 105 &  &  &  &  & InCdF$_{3}$ & 22 & 114 &  &  &  &  & \textcolor{blue}{TlHgF$_{3}$} & 17 & 33 &  &  &  &  & \textcolor{blue}{RbSrF$_{3}$} & 8 & \tabularnewline
\textcolor{blue}{PbTiO$_{3}$} & 190 &  &  &  &  &  & \textcolor{blue}{AgTaO$_{3}$} & 57 &  &  &  &  &  & ZnInF$_{3}$ & 22 & 32 &  &  &  &  & \textcolor{blue}{TlCdF$_{3}$} & 16 & 32 &  &  &  &  & \textcolor{black}{CsZnF$_{3}$} & 8 & 9\tabularnewline
\textcolor{red}{GaTaO$_{3}$} & 155 &  &  &  &  &  & \textcolor{red}{PbSiO$_{3}$} & 51 & 61 &  &  &  &  & ZnScF$_{3}$ & 22 & 30 &  &  &  &  & ZnYF$_{3}$ & 15 & 18 &  &  &  &  & \textcolor{blue}{CsSrF$_{3}$} & 8 & 12\tabularnewline
\textcolor{blue}{KNbO$_{3}$} & 122 & 835 &  &  &  &  & CdSbF$_{3}$ & 42 &  &  &  &  &  & InMgF$_{3}$ & 20 & 33 &  &  &  &  & \textcolor{black}{TlZnF$_{3}$} & 15 & 18 &  &  &  &  & \textcolor{blue}{KCaF$_{3}$} & 8 & 12\tabularnewline
\textcolor{blue}{NaNbO$_{3}$} & 116 &  &  &  &  &  & \textcolor{red}{TlSnF$_{3}$} & 35 & 136 &  &  &  &  & ZnAlF$_{3}$ & 20 & 23 &  &  &  &  & \textcolor{red}{NaBeF$_{3}$} & 14 & 25 &  &  &  &  & BaSiO$_{3}$ & 8 & \tabularnewline
SnSiO$_{3}$ & 116 & 443 &  &  &  &  & \textcolor{blue}{BaZrO$_{3}$} & 34 & 45 &  &  &  &  & ZnBiF$_{3}$ & 20 & 23 &  &  &  &  & \textcolor{red}{KPbF$_{3}$} & 14 &  &  &  &  &  & \textcolor{blue}{KCdF$_{3}$} & 8 & \tabularnewline
\textcolor{blue}{AgNbO$_{3}$} & 115 &  &  &  &  &  & \textcolor{blue}{SrHfO$_{3}$} & 31 &  &  &  &  &  & HgScF$_{3}$ & 19 & 23 &  &  &  &  & HgYF$_{3}$ & 13 & 15 &  &  &  &  & \textcolor{blue}{RbZnF$_{3}$} & 7 & 8\tabularnewline
\textcolor{blue}{PbHfO$_{3}$} & 111 &  &  &  &  &  & \textcolor{red}{RbSnF$_{3}$} & 30 & 217 &  &  &  &  & PdYF$_{3}$ & 19 &  &  &  &  &  & RbPbF$_{3}$ & 13 & 16 &  &  &  &  & \textcolor{blue}{RbCaF$_{3}$} & 7 & 10\tabularnewline
\textcolor{blue}{BaTiO$_{3}$} & 105 & 612 &  &  &  &  & GaZnF$_{3}$ & 29 & 95 &  &  &  &  & \textcolor{blue}{TlPbF$_{3}$} & 18 &  &  &  &  &  & \textcolor{black}{TlMgF$_{3}$} & 13 & 15 &  &  &  &  & \textcolor{blue}{CsCaF$_{3}$} & 7 & 8\tabularnewline
BeAlF$_{3}$ & 99 & 64 &  &  &  &  & BeYF$_{3}$ & 28 & 37 &  &  &  &  & \textcolor{blue}{SrSiO$_{3}$} & 18 & 19 &  &  &  &  & \textcolor{blue}{BaLiF$_{3}$} & 12 & 13 &  &  &  &  & \textcolor{blue}{KZnF$_{3}$} & 7 & 8\tabularnewline
AlMgF$_{3}$ & 97 &  &  &  &  &  & \textcolor{blue}{BaHfO$_{3}$} & 27 & 34 &  &  &  &  & CdYF$_{3}$ & 18 & 25 &  &  &  &  & \textcolor{blue}{AgZnF$_{3}$} & 10 & 11 &  &  &  &  & \textcolor{blue}{CsCdF$_{3}$} & 7 & 8\tabularnewline
\textcolor{red}{TlTaO$_{3}$} & 95 & 465 &  &  &  &  & \textcolor{red}{KSnF$_{3}$} & 26 &  &  &  &  &  & HgBiF$_{3}$ & 18 & 24 &  &  &  &  & CsBaF$_{3}$ & 10 &  &  &  &  &  & \textcolor{blue}{RbCdF$_{3}$} & 7 & 9\tabularnewline
\textcolor{blue}{SrTiO$_{3}$} & 81 & 221 &  &  &  &  & GaMgF$_{3}$ & 26 & 76 &  &  &  &  & \textcolor{blue}{CaSiO$_{3}$} & 18 &  &  &  &  &  & \textcolor{blue}{RbHgF$_{3}$} & 9 & 12 &  &  &  &  & \textcolor{blue}{KMgF$_{3}$} & 6 & 6\tabularnewline
\textcolor{blue}{RbTaO$_{3}$} & 78 & 249 &  &  &  &  & XeBiF$_{3}$ & 24 & 121 &  &  &  &  & HgInF$_{3}$ & 18 & 24 &  &  &  &  & \textcolor{black}{KHgF$_{3}$} & 9 &  &  &  &  &  & \textcolor{blue}{RbMgF$_{3}$} & 6 & 6\tabularnewline
AlZnF$_{3}$ & 63 &  &  &  &  &  & CdBiF$_{3}$ & 24 & 47 &  &  &  &  & XeScF$_{3}$ & 17 & 30 &  &  &  &  & \textcolor{blue}{CsHgF$_{3}$} & 9 & 11 &  &  &  &  &  &  & \tabularnewline
\textcolor{blue}{NaTaO$_{3}$} & 58 &  &  &  &  &  & InZnF$_{3}$ & 23 & 46 &  &  &  &  & TlCaF$_{3}$ & 17 & 39 &  &  &  &  & \textcolor{blue}{AgMgF$_{3}$} & 9 & 16 &  &  &  &  &  &  & \tabularnewline
\bottomrule
\end{tabular}
\par\end{centering}
\caption{List of the calculated relative static dielectric constants of cubic
perovskites that were found to be mechanically stable at 1000\,K
and at 300\,K. We highlight in blue the compounds that are experimentally
reported in the ideal cubic perovskite structure, and in red those
that are reported only in non-perovskite structures (same as in Ref.\,\citep{Ambroise_perovskites_kappa}).\label{tab:List-of-perovskites}}
\end{table*}

We list the stable compounds and their static dielectric constants
at 1000\,K, $\epsilon_{1000}$ -- and when they are also mechanically
stable at 300\,K, $\epsilon_{300}$ -- in Table \ref{tab:List-of-perovskites}.
We highlight in blue 35 perovskites that have been reported experimentally
in the ideal cubic structure, and in red 9 compounds that are reported
only in a non-perovskite form (see References in Ref.\,\citep{Ambroise_perovskites_kappa}).
On top of the list of the reliable materials, we see that niobates,
and in particular KNbO$_{3}$, appear to be the most promising compounds
for high-temperature applications, along with PbTiO$_{3}$. Those
compounds are actually already used at an industrial scale for this
property.

\begin{figure*}
\begin{centering}
\includegraphics[height=5cm]{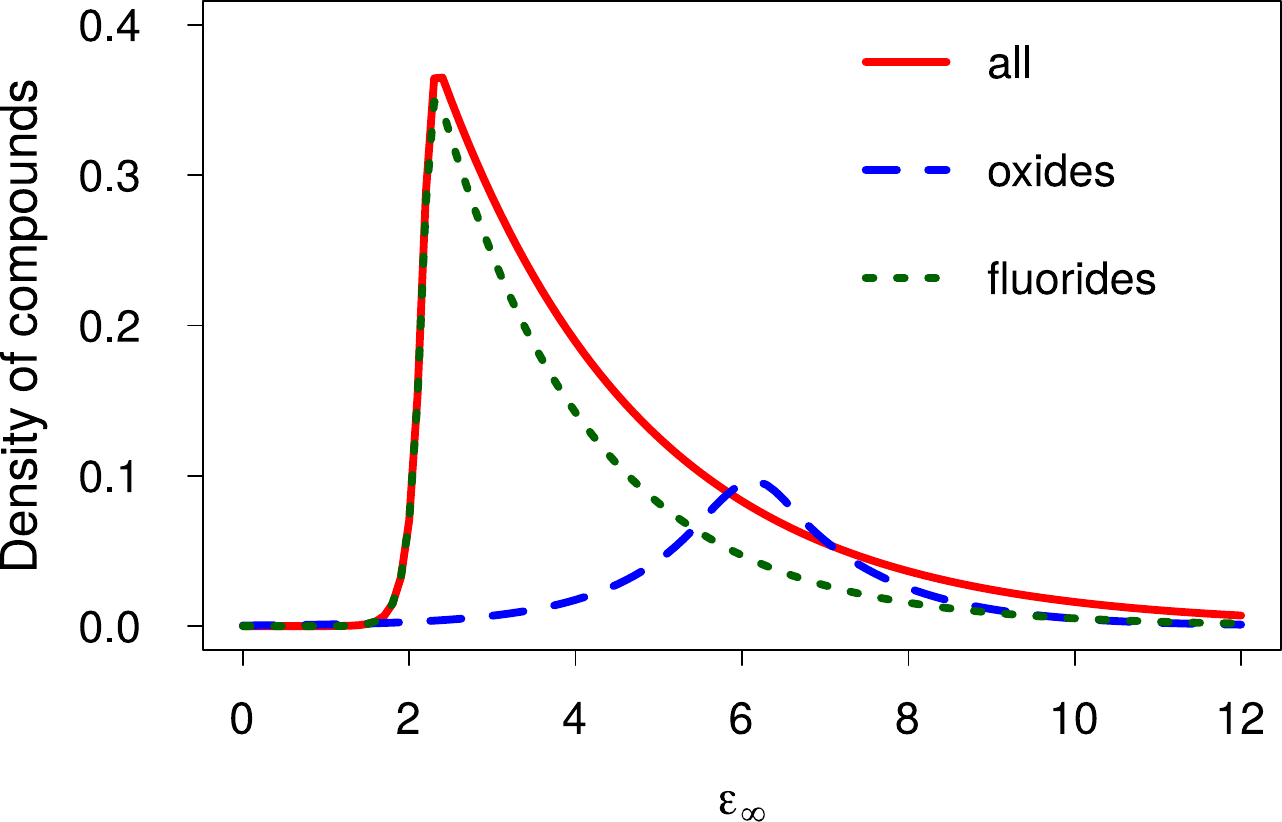}\hspace{1cm}\includegraphics[height=5cm]{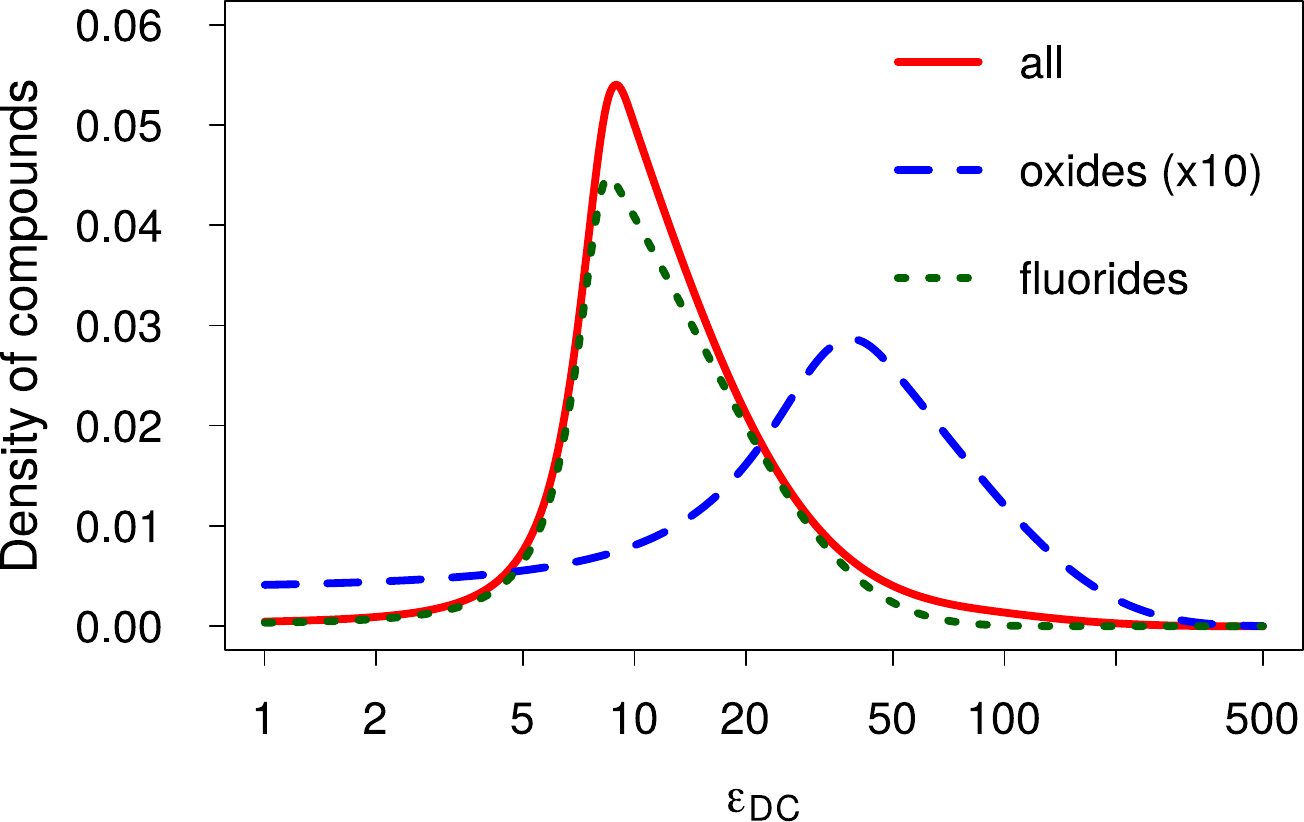}
\par\end{centering}
\caption{Distribution of compounds as a function of the static dielectric constant
including only the electronic contribution ($\epsilon_{\infty}$,
left) and also the lattice contribution at 1000\,K ($\epsilon_{DC}$,
right). The red curve corresponds to the distribution for all mechanically
stable compounds. The green curve corresponds to the distribution
for fluorides only. The blue curve corresponds to the distribution
for oxides only (the scale has been multiplied by 10 on the right
graph).\label{fig:Fluorides_vs_oxides}}
\end{figure*}

\begin{figure*}
\begin{centering}
\includegraphics[width=8.5cm]{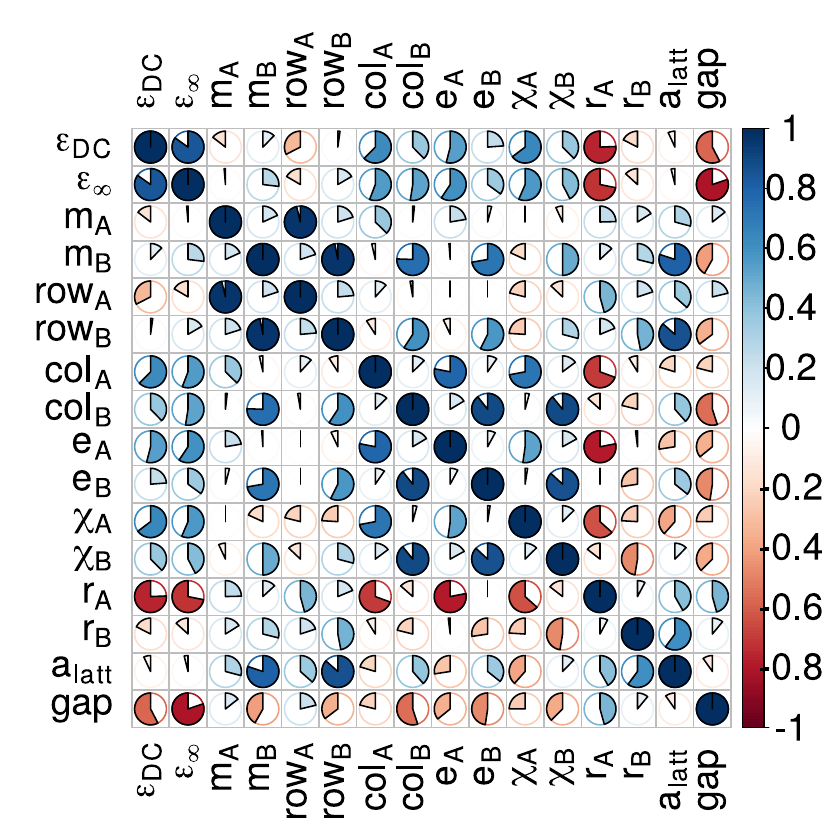}\hspace{0.5cm}\includegraphics[width=8.5cm]{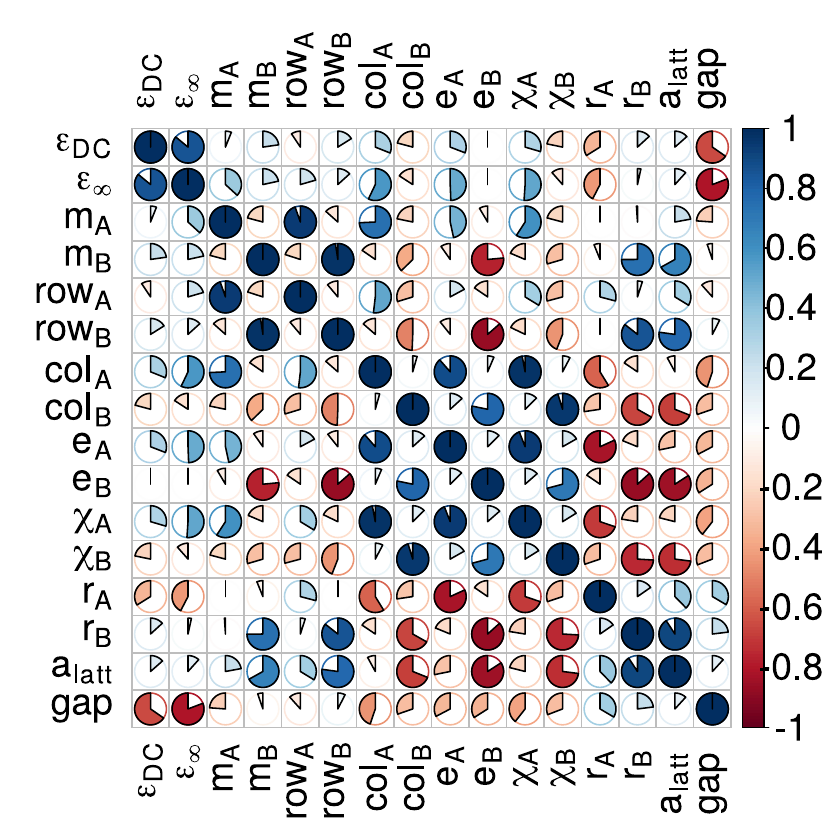}
\par\end{centering}
\caption{Spearman correlograms between the static dielectric constant at 1000\,K
$\epsilon_{DC}$, the infinite frequency (static electronic) dielectric
constant $\epsilon_{\infty}$, the masses m\protect\textsubscript{\textit{A}}
and m\protect\textsubscript{\textit{B}} of atoms at sites \textit{A}
and \textit{B} of the perovskite \textit{ABX}$_{3}$, their position
in the periodic table, their electronegativity e\protect\textsubscript{\textit{A}},
e\protect\textsubscript{\textit{B}}, their Pettifor scale \textgreek{q}\protect\textsubscript{\textit{A}},
\textgreek{q}\protect\textsubscript{\textit{B}}, their ionic radius
r\protect\textsubscript{\textit{A}}, r\protect\textsubscript{\textit{B}},
the lattice parameter of the compound a\protect\textsubscript{latt}
and its electronic band gap, for mechanically stable fluorides (left)
and oxides (right).\label{fig:Correlograms}}
\end{figure*}

Figure \ref{fig:Fluorides_vs_oxides} illustrates the distribution
of compounds as a function of the dielectric constant, both for the
ion-clamped (with electronic contribution only) and for the static
one at 1000\,K, based on logspline density estimation as implemented in R \citep{logspline}.
As can be seen already in Table \ref{tab:List-of-perovskites},
the effect of the lattice screening is sometimes much stronger in oxides,
and as a consequence there is a larger dispersion of the values of
the static dielectric constant than for the fluorides. We also remark
that the high-frequency dielectric properties reflect the fact that
oxygen is more polarizable than fluorine, and this
stronger polarizability might be linked to the stronger reaction
of the lattice. 

We delve into the question by computing the Spearman correlograms of
the static dielectric constant at 1000\,K $\epsilon_{DC}$ with $\epsilon_{\infty}$,
and with several simple descriptors of the cubic perovskite structure
\textit{ABX}$_{3}$: the masses of atoms \textit{A} and \textit{B},
their position in the periodic table, their electronegativity and
Pettifor scales, their ionic radii, the lattice parameter of the compound
and its electronic band gap. This is shown in Fig. \ref{fig:Correlograms},
for fluorides and for oxides separately. We indeed find out that there
is a strong correlation between the high-frequency and the static
dielectric constants, which was not obvious since the transition temperature
plays a crucial role, even if $\epsilon_{\infty}$ is a factor in
Eq. (\ref{eq:LST}). We note that we have obtained similar correlograms
when we computed the dielectric constants using the phonon spectra
with a different exchange-correlation functional from our previous
work \citep{Ambroise_perovskites_kappa}.

Interestingly, the ionic radius of atom \textit{A}
and the electronic band gap have significant negative correlation with
the static dielectric constant. This has indeed been observed empirically
based on experimental data \citep{Robertson_high_dielectric_constant}.
Since the stability of the cubic phase is known to be intimately linked
with geometric factors \citep{Goldschmidt_tolerance_factor_1926},
it is likely that the ionic radius of atom \textit{A} influences both
the electronic polarizability of the compound and its transition temperature.
A consequence of this finding is that if one wishes only to know the
order of the static dielectric constant between perovskites in the
same cubic phase, the high-frequency limit is actually a sufficiently
good indication. This is valid for instance for the series studied
above, BaTiO$_{3}$, SrTiO$_{3}$ and KTaO$_{3}$. It also explains
the good Spearman correlation that was found in Ref.\,\citep{Petousis_HT_dielectric_constant}
between calculated values using DFPT at the ground state and experimental
measurements at room temperature, despite the occasional strong discrepancies
in the order of magnitude for perovskite compounds.

\section{Conclusion}

In conclusion, we have shown that the temperature-dependence of the
static dielectric constant can be calculated from first principles
with reasonable agreement with experiment over a large temperature
range, simply taking into account the temperature-dependence of the
phonon spectrum and ground-state properties. In particular, we have
evaluated quantitatively the effects of anharmonicity on the ion-clamped
dielectric constant and Born charges, as well as the inclusion of
lifetime effects in the generalized Lyddane-Sachs-Teller relations.
Then, we have conducted a high-throughput study of the high-temperature
static dielectric constant in oxide and fluoride cubic perovskites.
We observed that the rank of the static dielectric constant at high
temperature is already well described by the ion-clamped value, and
mostly linked to the ionic radius of atom \textit{A} of the perovskite
\textit{ABX}$_{3}$. These findings bring coherence to the puzzling
correlations between ground-state calculations and temperature-dependent
experimental measurements that have been found in previous high-throughput
studies of the dielectric constant. We hope that our study will motivate
high-throughput computational search of dielectric materials for specific
applications in harsh temperature conditions.
\begin{acknowledgments}
This work was supported by the Carnot project SIEVE and HPC resources
from GENCI-TGCC (projects A0030910242 and A0060910242). SC acknowledges
financial support by DOD-ONR (N00014-20-1-2525 and N00014-20-1-2200)
and the Alexander von Humboldt-Foundation.
\end{acknowledgments}

\input{Dielectric_constant_resubmission.bbl}

\end{document}

%% file: Dielectric_constant_resubmission.bbl
%